\documentclass[12pt]{article}
\usepackage{pdfpages}
\usepackage{url}
\def\beq{\begin{equation}}
\def\eeq#1{\label{#1}\end{equation}}
\def\eeqn{\end{equation}}

\def\beqa{\begin{eqnarray}}
\def\eeqa#1{\label{#1}\end{eqnarray}}
\def\eeqan{\end{eqnarray}}

\let\bar=\overbar

\def\Dslash{\not{\hbox{\kern-4pt $D$}}}
\def\dslash{\not{\hbox{\kern-2pt $\del$}}}

\def\msb{{\bar{\ssstyle M \kern -1pt S}}}

\def\Title#1{\begin{center} {\Large {\bf #1} } \end{center}}

\begin{document}

\Title{The ATLAS Trigger Algorithms Upgrade and Performance in Run-2}

\bigskip\bigskip


\begin{raggedright}  
{\it Dr. Catrin Bernius, on behalf of the ATLAS Collaboration\index{Reggiano, D.}\\
SLAC National Accelerator Laboratory \\
2575 Sand Hill Road, Menlo Park, CA 94025\\}
\bigskip

Talk presented at the APS Division of Particles and Fields Meeting (DPF 2017), July 31-August 4, 2017, Fermilab. C170731
\bigskip\bigskip
\end{raggedright}

\section{Introduction}
The trigger system~\cite{trigger} of the ATLAS experiment~\cite{atlas} at the Large Hadron Collider (LHC) was efficiently operated at instantaneous luminosities up to $\mathrm{8 \times 10^{33}}$ cm$\mathrm{^{-2}}$ s${^{-1}}$ and at centre-of-mass energies of $\mathrm{\sqrt{s}}$ = 7 TeV and 8 TeV during Run-1 (2009-2012).
In preparation for Run-2 (2015-2018) substantial upgrades and modifications to the trigger system and the software and algorithms to select events of interest were carried out. 
These were necessary as the trigger rates were expected to increase due to the increases beyond the instantaneous design luminosity of 10$^{34}$cm${^{-2}} $ s${^{-1}}$, the number of proton-proton collisions per bunch-crossing (pile-up), and the $\mathrm{\sqrt{s}}$ to 13 TeV.
The first two years of Run-2 have presented a challenging environment, but the ATLAS trigger system has continued to operate efficiently and reliably. While the centre-of-mass energy will remain the same for the remaining two years of Run-2, the instantaneous luminosity is expected to increase up to $\mathrm{1.7 \times 10^{34}}$ cm$\mathrm{^{-2}}$ s${^{-1}}$ together with a further increase in pile-up. To continue the stable operation of the trigger system~\cite{Run2TriggerPerformance}, the trigger strategies were substantially improved with increased rejection power during 2017 data taking. 
This document summarises the design of the ATLAS trigger system and presents a selection of just a few of the many improvements to the trigger selection algorithms.

\section{The Run-2 ATLAS Trigger System}
The Run-2 ATLAS Trigger and Data Acquisition (TDAQ) system consists of a hardware-based first level (Level-1) and a software-based high level trigger (HLT).  
Level-1 runs with a fixed latency of 2.5 $\mu$s and reduces the event rate from 40 MHz to 100 kHz. The Level-1 trigger decision is formed by the Central Trigger Processor, which receives information from the Level-1 calorimeter (Level-1 Calo) and the Level-1 muon (Level-1 Muon) triggers as well as from the topological trigger (L1Topo) which performs selections based on geometric or kinematic associations between trigger objects received from Level-1 Calo and/or Level-1 Muon. \\
The HLT receives the input from Level-1 in the form of geometrical regions in $\eta$ and $\phi$, so-called regions of interest (RoI). It runs on a single farm of about 40000 processors, that accept or reject the event within 300 ms, and performs object reconstruction in the RoI using algorithms that are as close as possible to their counterparts used in offline reconstruction. The Fast Tracker (FTK)~\cite{FTK} which is currently being installed and commissioned will provide hardware-based tracking to the HLT.

\section{Algorithm Upgrades and Performance}
Events are selected based on physics signatures such as the presence of energetic leptons, photons, jets or large missing energy. Each one of these signatures is reconstructed and identified using different strategies. While there have been many improvements in the reconstruction and identification across all signatures, in the following selected improvements in the electron, muon, jet and missing transverse energy signature group will be described. A lot of work and effort has gone into improvements in the $b$-jet and $B$-physics trigger signatures which will not be discussed here. 

\subsection{Electrons}
\begin{figure}[t]
	\begin{center}
		\includegraphics[width=0.49\textwidth]{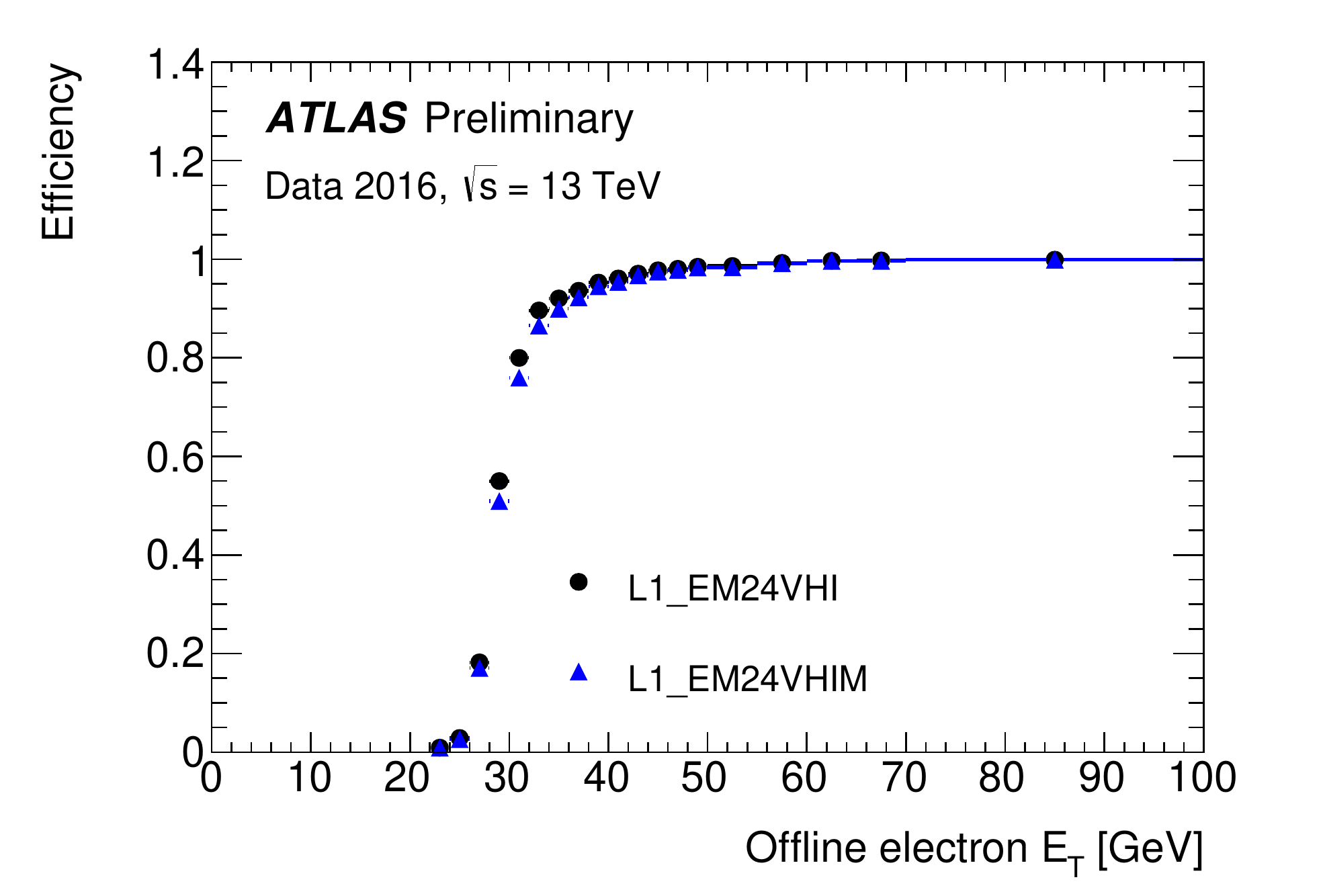}
		\includegraphics[width=0.46\textwidth]{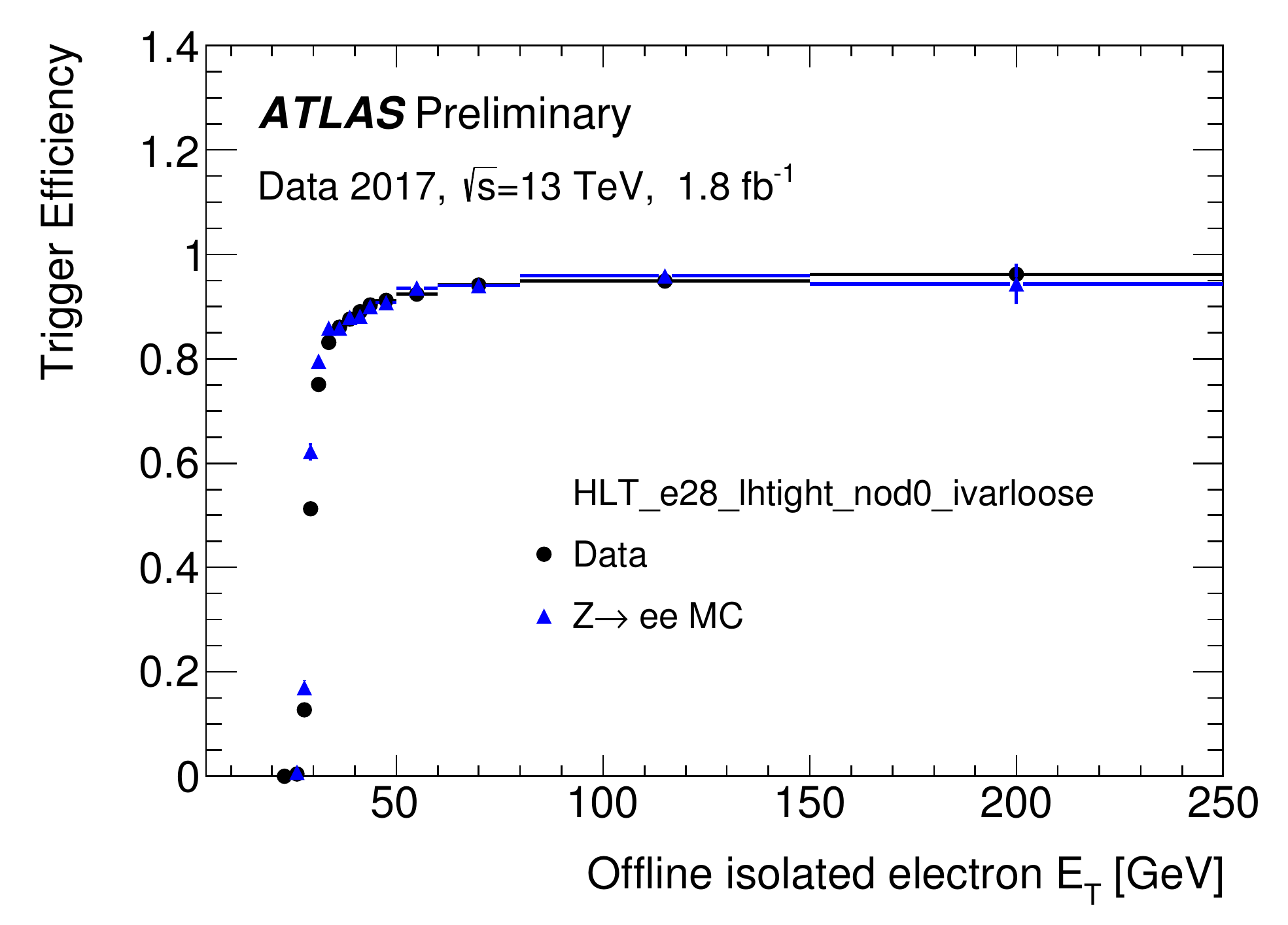}
		\caption{Efficiency of the Level-1 (left) and of the HLT triggers (right) as a function of the offline electron candidate $E_{\mathrm{T}}$. The Level-1 triggers require an isolated electromagnetic (EM) cluster with $E_{\mathrm{T}} >$ 24 GeV. ``I" (``IM") stands for the isolation applied for EM clusters with $E_{\mathrm{T}} <$ 50 GeV where the $E_{\mathrm{T}}$ in an annulus of calorimeter towers around the EM candidate relative to the EM cluster $E_{\mathrm{T}}$ is required to be less than max$\{$2 GeV, $E_{\mathrm{T}}$/8$-$1.8 GeV$\}$ (max$\{$2 GeV, $E_{\mathrm{T}}$/8$-$2.0 GeV$\}$). The HLT trigger requires an electron candidate with $E_{\mathrm{T}} >$ 28 GeV satisfying the likelihood-based tight identification without applying transverse impact parameter requirements but applying variable-size cone isolation~\cite{electron_publicresults}. }
		\label{Fig:electron}
	\end{center}
\end{figure}

Many improvements were made in the electron and photon trigger signature. Highlighted here is the implementation of new Level-1 electromagnetic (EM) medium isolation cuts to reduce the rate of the lowest unprescaled Level-1 triggers while keeping the efficiency loss as low as possible in order to cope with the increasing luminosity in 2017. Reducing the trigger rate through isolation means that the single electron trigger thresholds can be kept low. Figure~\ref{Fig:electron} (left) shows a comparison of the efficiency curves of the new isolation cuts (L1\_EM24VHIM) with the default cuts (L1\_EM24VHI) used for 2016 data taking.
Figure~\ref{Fig:electron} (right) shows the efficiency of a single electron trigger at the HLT as a function of the transverse energy ($E_{\mathrm{T}}$) of the offline electron candidate in comparison with Monte Carlo simulation, showing good agreement and a sharp turn-on curve.  

\subsection{Muons}
Until recently, the muon stand-alone trigger algorithm (FastMuonSA) which determines the transverse momentum, $p_\mathrm{T}$, had not included information from the ATLAS muon Cathode Strip Chamber (CSC) system, located in the forward region of the ATLAS detector between $2.0 < |\eta| < 2.7$. In 2017 it has been shown that including hits from the CSCs results in an improvement of the $1/p_\mathrm{T,offline}$ resolution in all $p_\mathrm{T}$ ranges. 
\begin{figure}[tb]
	\begin{center}
		\includegraphics[width=0.75\textwidth]{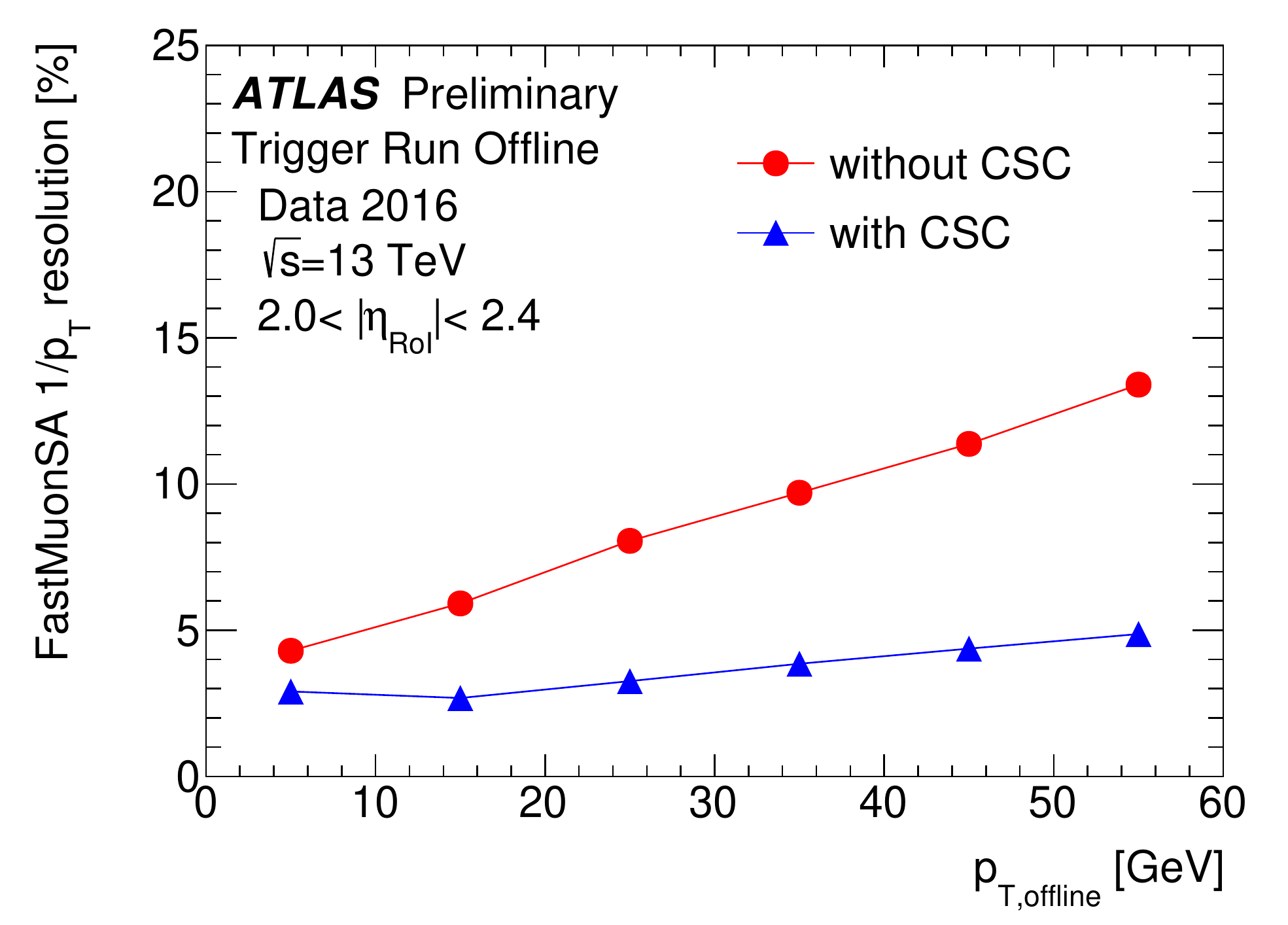}
		\caption{The resolution of the inverse $p_\mathrm{T}$ from the fast muon stand-alone trigger algorithm (FastMuonSA) is shown as a function of the offline muon $p_\mathrm{T}$. The blue triangles show the resolution when FastMuonSA uses hits from CSC chambers, and the red circles show that when FastMuonSA does not use hits from CSC chambers. The results were obtained by rerunning FastMuonSA on the 2016 data~\cite{muon_publicresults}. }
		\label{Fig:muon_res}
	\end{center}
\end{figure}
\begin{figure}
	\begin{center}
		\includegraphics[width=1.0\textwidth]{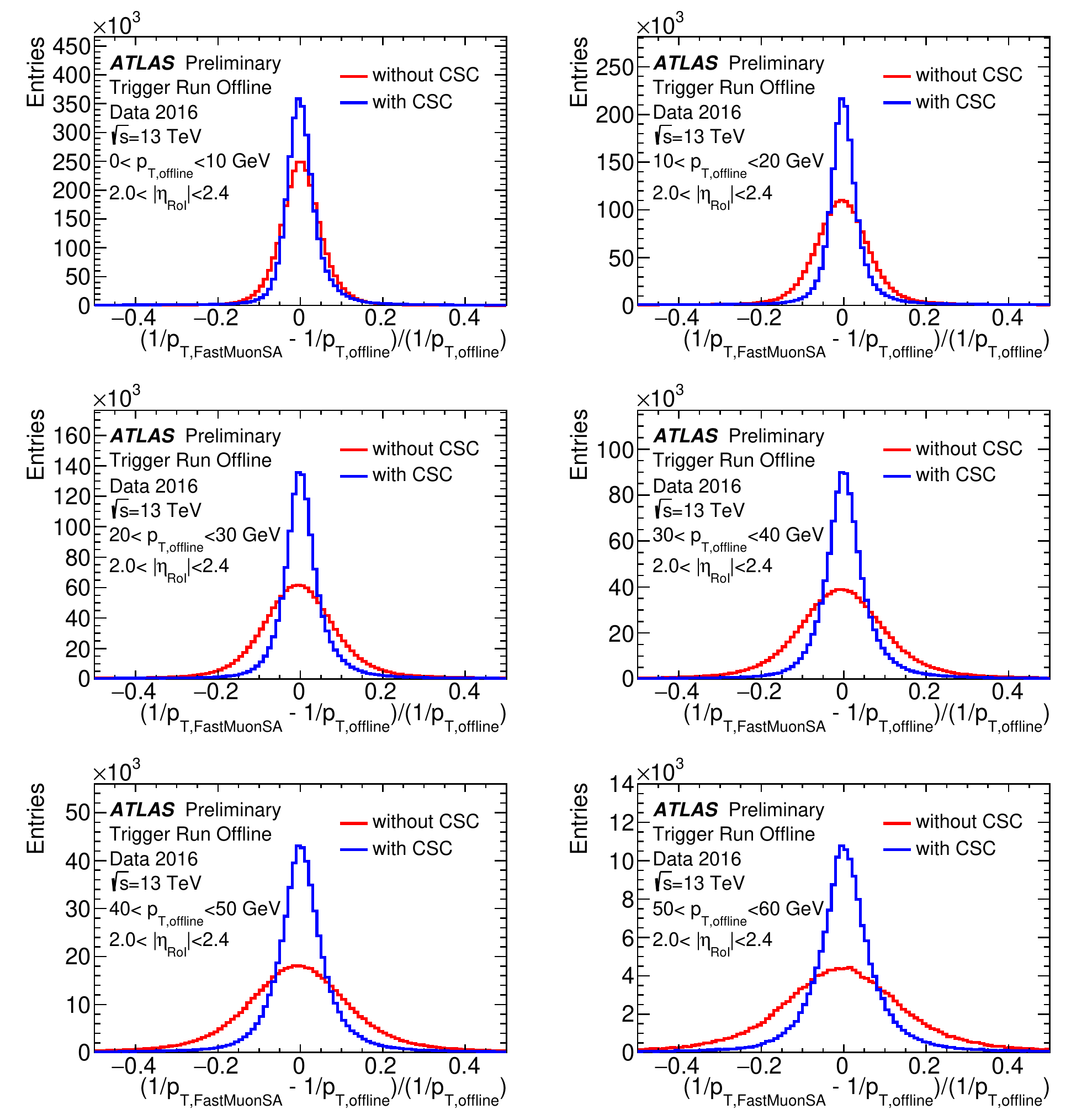}
		\caption{Fractional residual of inverse $p_\mathrm{T}$ is shown where $p_\mathrm{T,FastMuonSA}$ is the $p_\mathrm{T}$ reconstructed by the fast muon stand-alone trigger algorithm (FastMuonSA) and $p_\mathrm{T,offline}$ is the $p_\mathrm{T}$ given by the offline reconstruction~\cite{muon_publicresults}. }
		\label{Fig:muon_allres}
	\end{center}
\end{figure}
Figure~\ref{Fig:muon_res} shows the inverse $p_\mathrm{T}$ resolution as a function of the offline muon $p_\mathrm{T}$ with and without using hits from the CSC chambers. The resolution is extracted by taking the $\sigma$ of a Gaussian fit to the distribution of the fractional residual of inverse $p_\mathrm{T}$, i.e. ($1/p_\mathrm{T,FastMuonSA} - 1/p_\mathrm{T,offline})/(1/p_\mathrm{T,offline})$.  
The fractional residual of the inverse $p_\mathrm{T,offline}$ for all $\eta$ ranges is shown in Figure~\ref{Fig:muon_allres}.

\subsection{Taus}
Various improvements and changes were made to e.g. the online tau energy scale corrections and online tau identification using a Boosted Decision Tree to reflect changes made to the offline reconstruction methods. Additionally, L1Topo trigger items are now used for triggering on di-$\tau$ and lepton + $\tau$ signatures. 

\subsection{Jets}
\begin{figure}
	\begin{center}
		\includegraphics[width=0.7\textwidth]{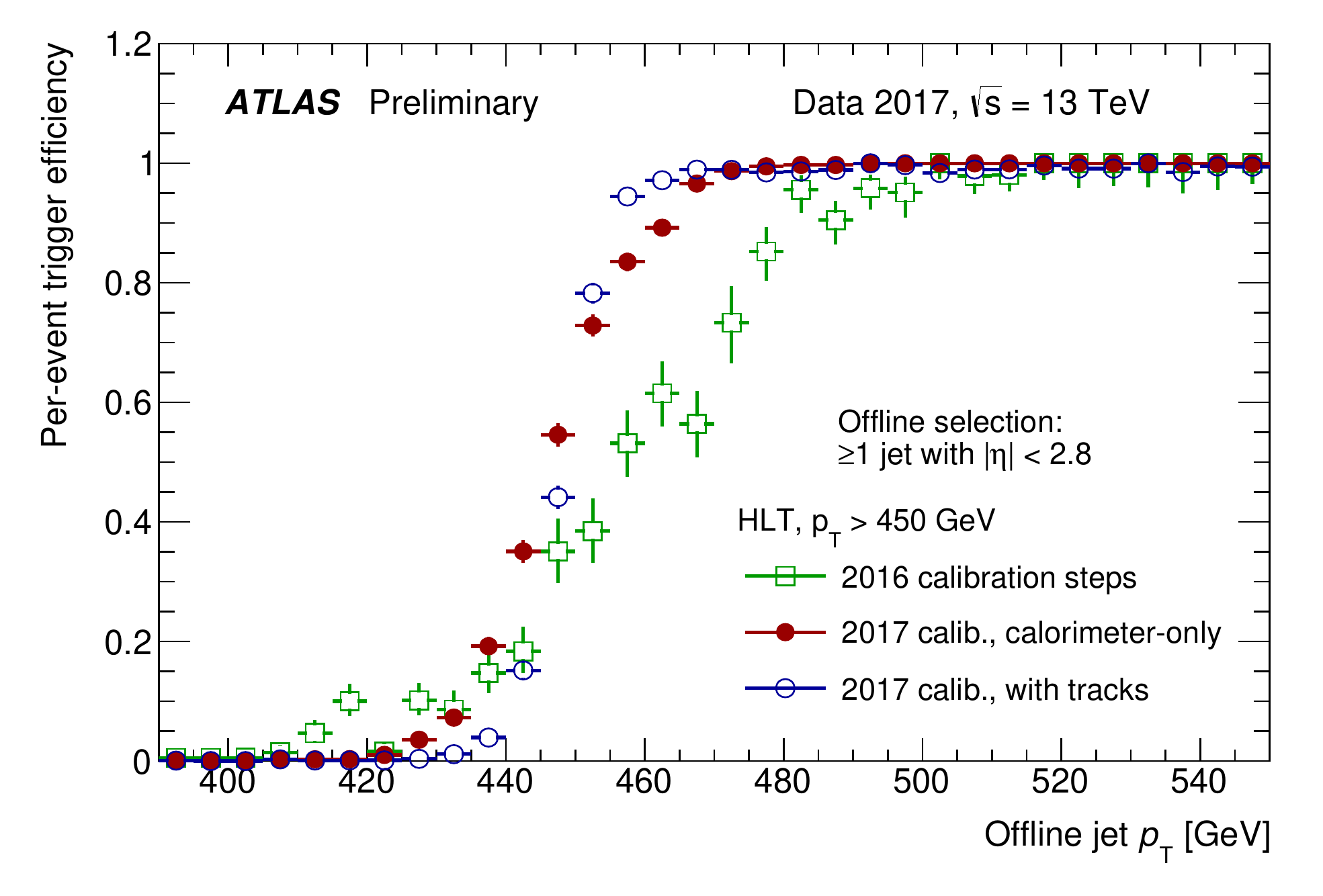}
		\caption{Efficiencies are shown for an unprescaled single-jet trigger with three different calibrations applied to jets in the ATLAS high level trigger (HLT). Offline jets are selected with $|\eta| < 2.8$. The red closed circles show the calibration steps applied in 2016 data, the blue open circles show the updated calibration applied in 2017, utilising only calorimeter information, and the green open squares show the updated calibration additionally with track information~\cite{jet_publicresults}. }
		\label{Fig:jet_smallR}
	\end{center}
\end{figure}

To determine the jet energy scale, its uncertainty, and to achieve an optimal jet energy resolution, several calibration schemes have been developed in ATLAS. The global sequential calibration (GSC)~\cite{gsc} corrects jets according to global jet observables such as the longitudinal structure of the energy depositions within the calorimeters, tracking information associated to the jet, and information related to the activity in the muon chambers behind a jet without changing the overall energy scale. They can be split into parts involving calorimeter-based variables, and parts involving track-based variables. Since tracking is not guaranteed to be available for all jet thresholds, options are provided with and without the track-based corrections. The data-driven $\eta$-intercalibration correction~\cite{insitu} is the most important in situ correction added, and fixes differences in jet response as a function of $\eta$. 
Figure~\ref{Fig:jet_smallR} compares the efficiencies for a single jet trigger for the calibration applied in 2016, the updated calibration applied in 2017 using only calorimeter information and the updated calibration additionally with track information. 
The additional corrections applied to the jets reconstructed at the HLT allow for improved agreement between the scale of trigger and offline jets as a function of both $\eta$ and $p_\mathrm{T}$, and thus the trigger efficiency rises much more rapidly.
\begin{figure}
	\begin{center}
		\includegraphics[width=0.49\textwidth]{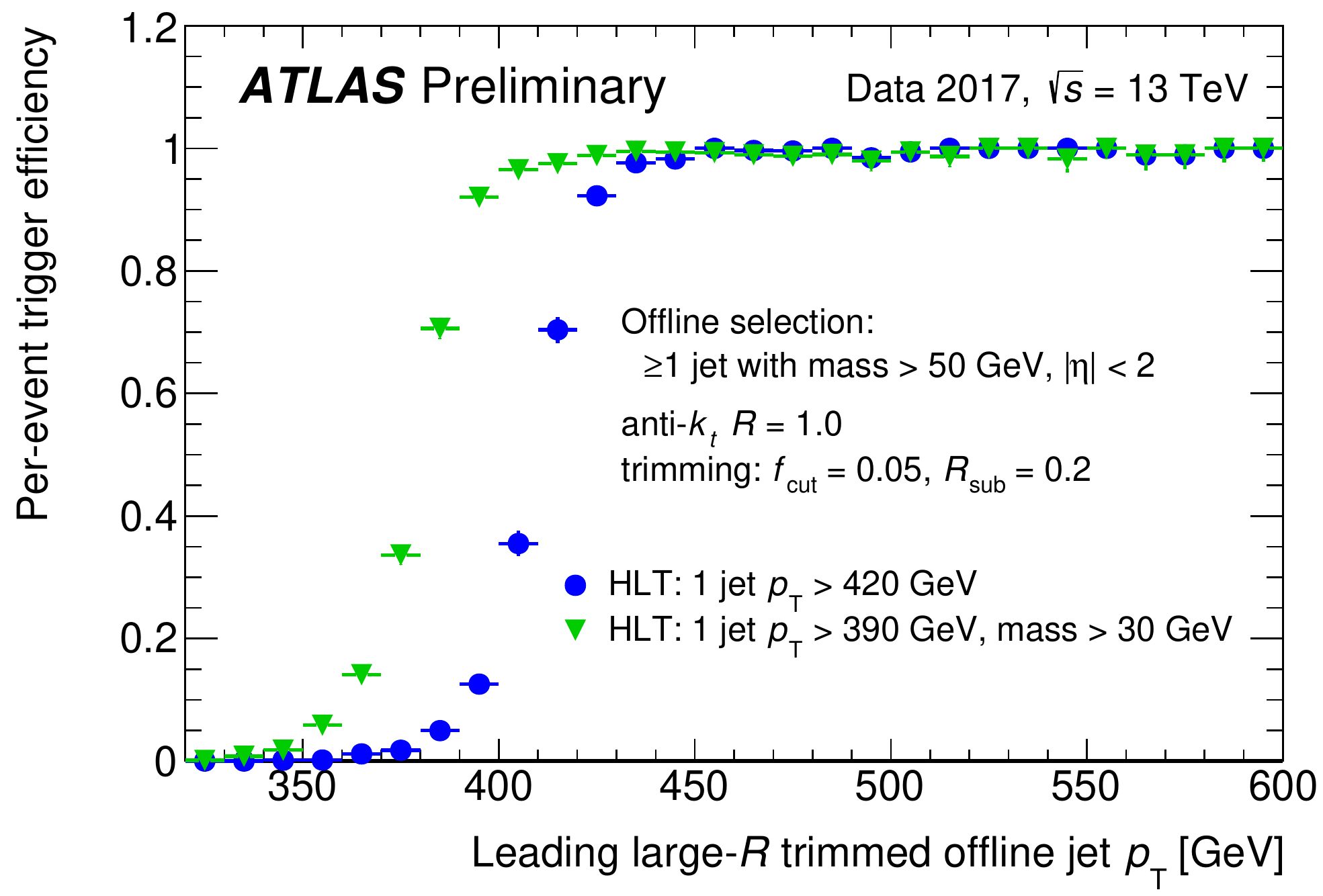}
		\includegraphics[width=0.49\textwidth]{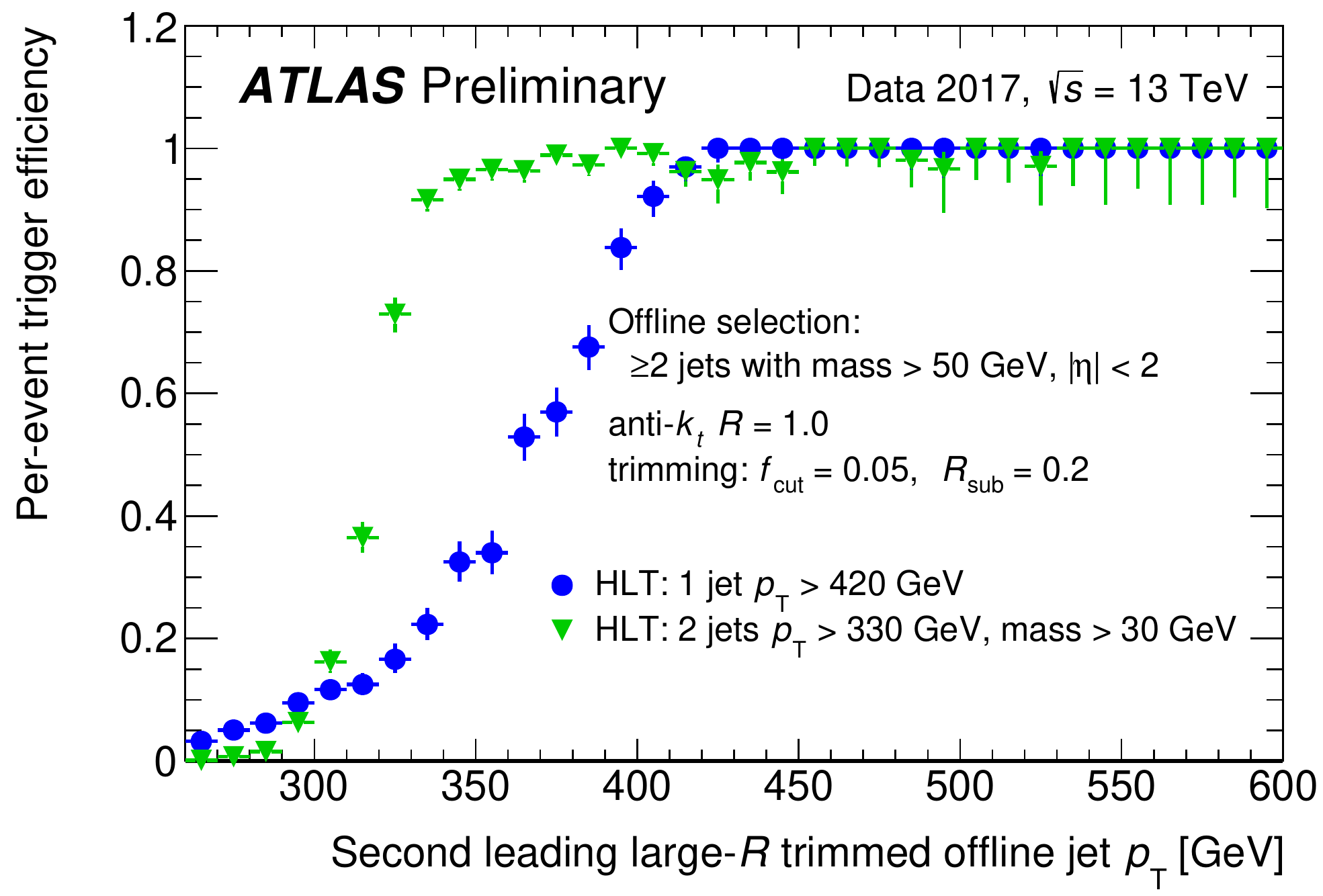}
		\caption{Efficiencies for HLT large-$R$ single-jet triggers are shown as a function of the leading (left) and second leading (right) offline trimmed jet $p_\mathrm{T}$ for jets with $|\eta| < 2.0$ and jet mass above 50 GeV. Two large-$R$ jet triggers, from the 2017 menu, are shown. Blue circles represent a trimmed large-$R$ jet trigger with a $p_\mathrm{T}$ threshold of 420 GeV. Adding an additional 30 GeV cut on the jet mass of the selected trimmed trigger jet is shown by green triangles~\cite{jet_publicresults}. }
		\label{Fig:jet_largeR}
	\end{center}
\end{figure}
Large jets, i.e. reconstructed with the anti-k$_{\mathrm{T}}$ algorithm with a radius parameter $R$ = 1, are more susceptible to pile-up. Applying offline grooming techniques like trimming~\cite{trimming} can help to reduce the effect of pile-up by removing soft and wide angle radiation from the jet clustering history which significantly pushes the mass distribtuion for light jets towards zero while having only a minimal effect on jets from heavy particle decays. 
In 2017, trimming has been implemented in large-$R$ jet triggers at the HLT. Jet constituents are re-clustered with $R$ = 0.2 to form sub-jets, and are removed if $p_{\mathrm{T, sub-jet}}/p_{\mathrm{T, jet}} < 4$ $\%$. This is slightly altered from what is applied offline (5 $\%$) to recover efficiency losses. 
As trimming produces a stable distribution of the jet mass as a function of pile-up, applying a mass cut significantly suppresses the QCD di-jet background, allowing a lower $p_\mathrm{T}$ threshold, while retaining nearly all signal-like jets. 
This is shown in Figure~\ref{Fig:jet_largeR} which compares the efficiencies for triggers with and without applying a mass cut for large-$R$ single jet triggers at the HLT as a function of the leading (left) and second-leading (right) jet $p_{\mathrm{T}}$. 

\subsection{Missing Transverse Energy}
\begin{figure}
	\begin{center}
		\includegraphics[width=0.5\textwidth]{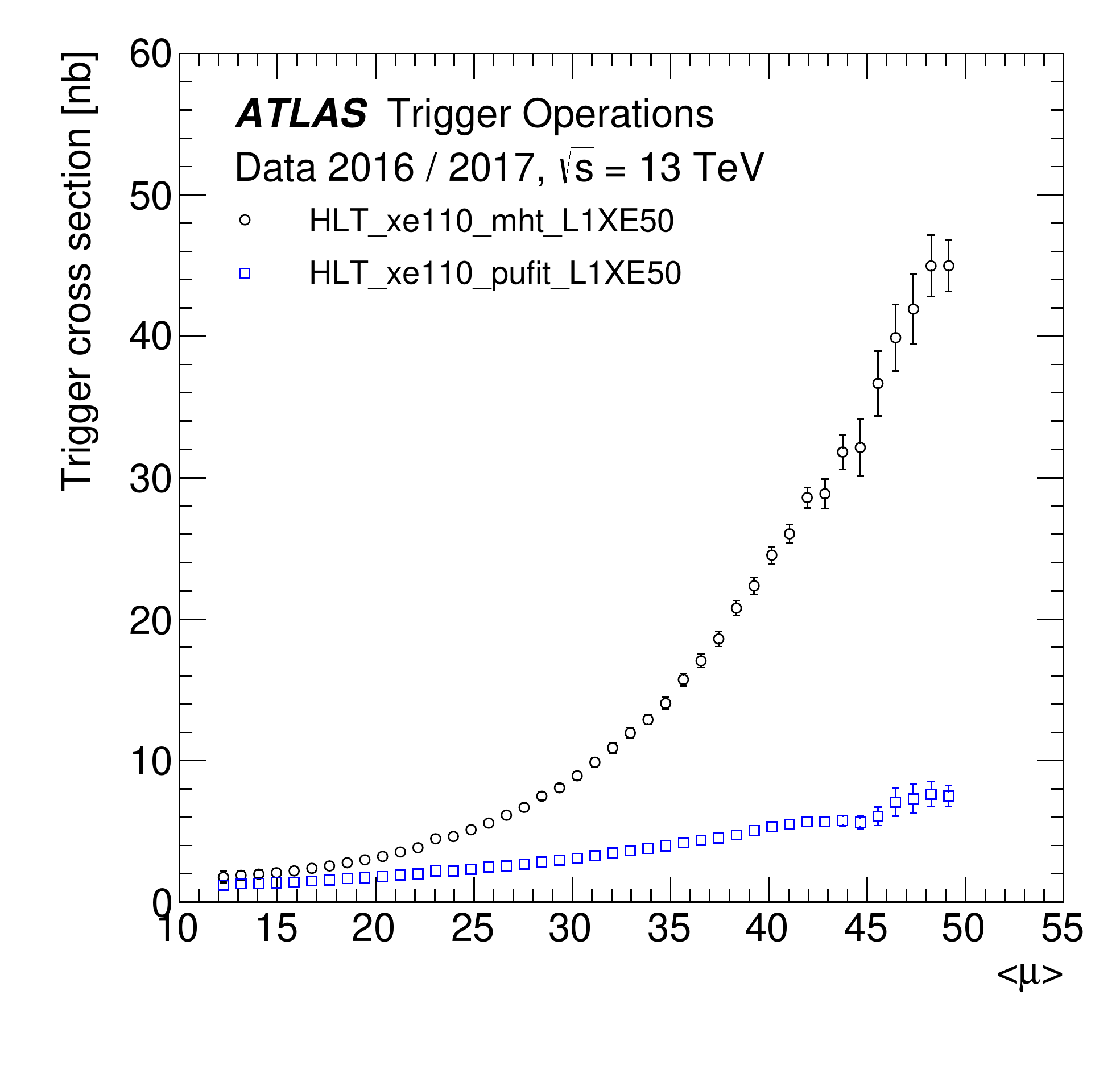}
		\caption{The trigger cross-section as measured by using online rate and luminosity is compared for the main trigger $E\mathrm{_T^{miss}}$ reconstruction algorithm used in 2016 \textit{mht} and 2017 \textit{PUFit} as a function of the mean number of simultaneous interactions per proton-proton bunch crossing averaged over all bunches circulating in the LHC. The triggers HLT$\_$xe110$\_$mht$\_$L1XE50 and HLT$\_$xe110$\_$pufit$\_$L1XE50 are used as representative benchmarks of the 2016 and 2017 data-taking campaigns, respectively~\cite{met_pubresults}.}
		\label{Fig:met_rate}
	\end{center}
\end{figure}

\begin{figure}
	\begin{center}
		\includegraphics[width=0.49\textwidth]{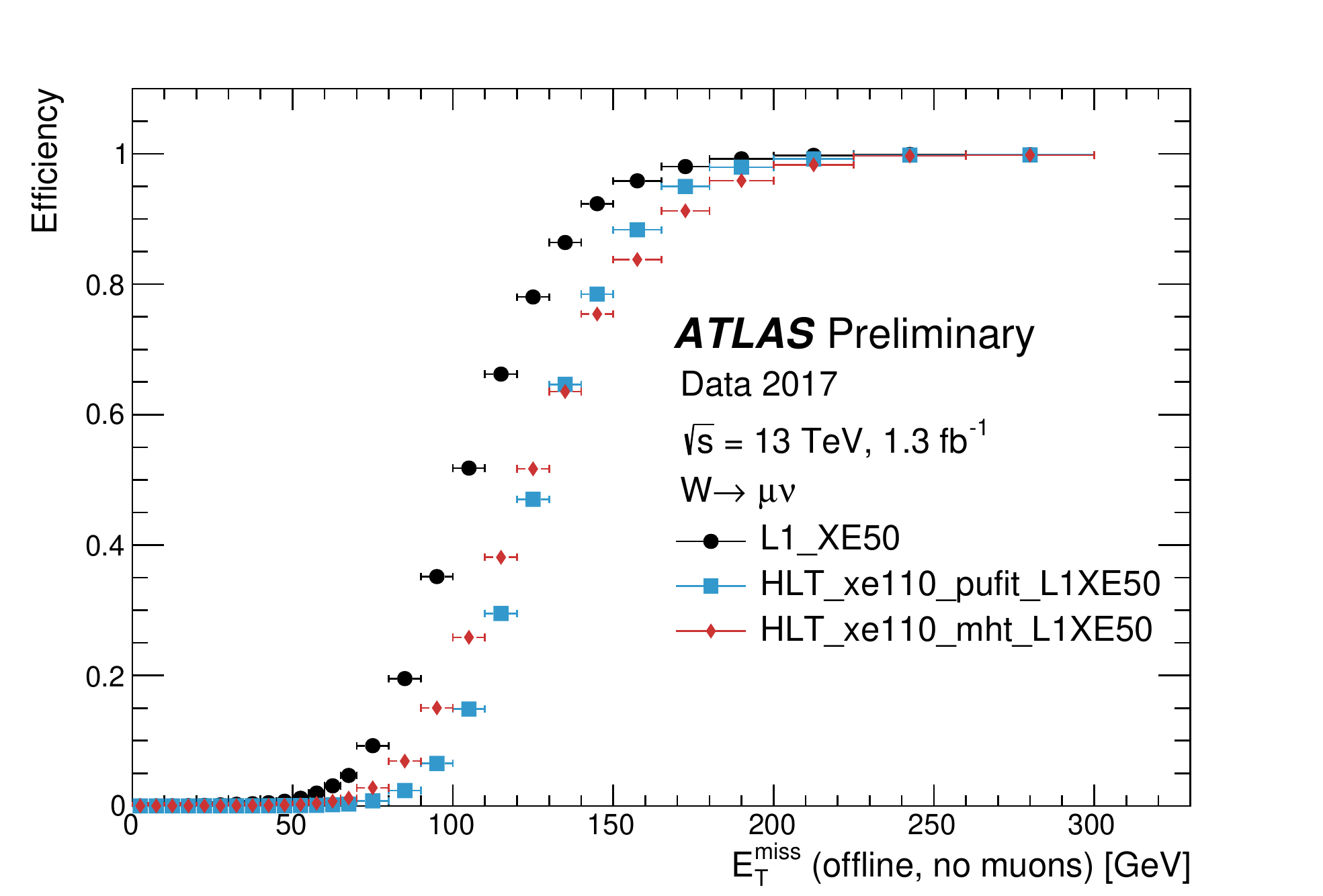}
		\includegraphics[width=0.49\textwidth]{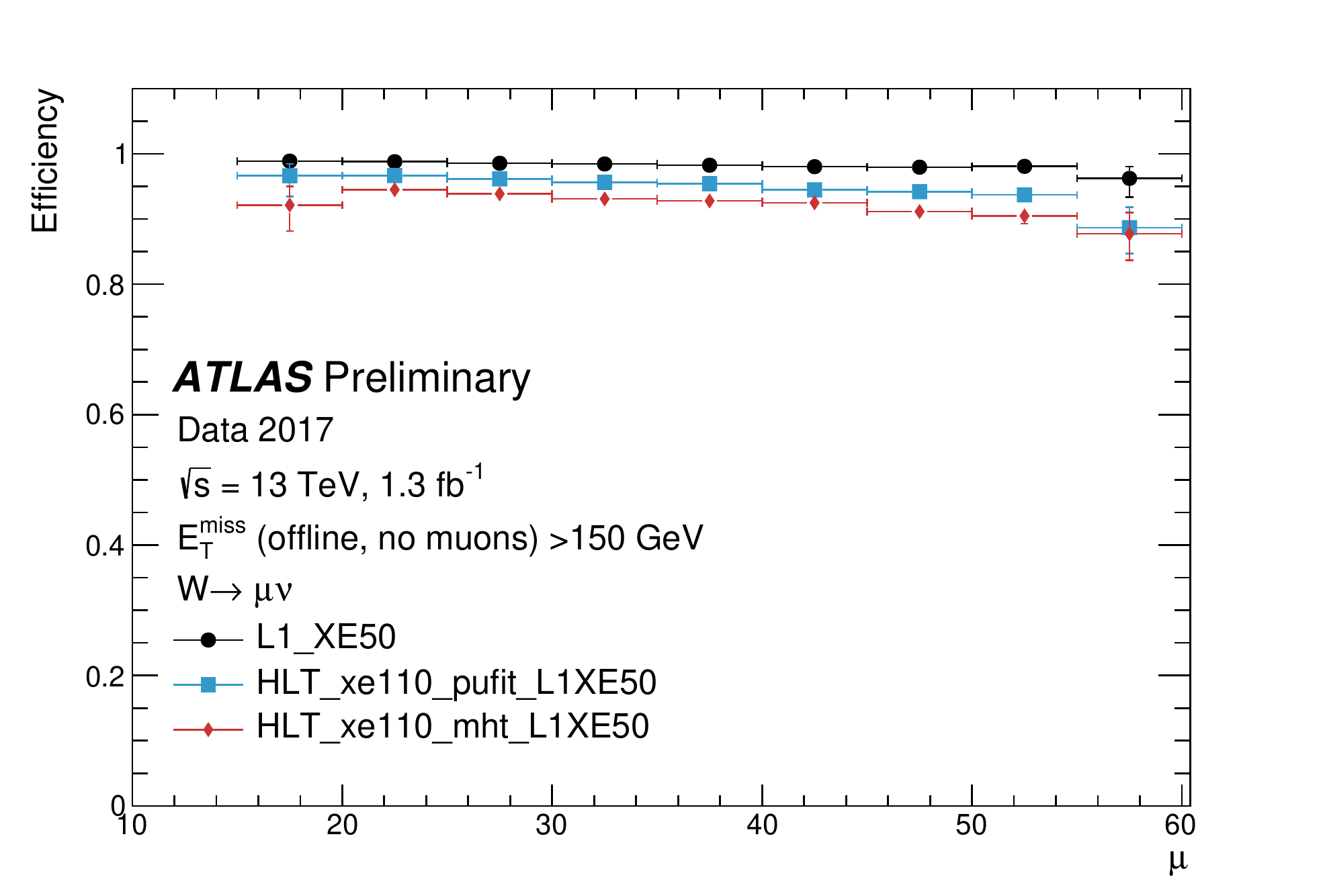}
		\caption{The combined L1 and HLT efficiency of the missing transverse energy triggers HLT$\_$xe110$\_$mht$\_$L1XE50 and HLT$\_$xe110$\_$pufit$\_$L1XE50 as well as the efficiency of the corresponding L1 trigger (L1$\_$XE50) are shown as a function of the reconstructed $E\mathrm{_T^{miss}}$ (modified to count muons as invisible) (left) and as a function of the mean number of simultaneous interactions in a given proton-proton bunch crossing (pile-up) (right)~\cite{met_pubresults}. }
		\label{Fig:met_eff}
	\end{center}
\end{figure}

The missing transverse energy ($E\mathrm{_T^{miss}}$) trigger rates are severely affected by detector noise, mis-measurements and increase non-linearly with the number of pile-up interactions. In 2016, the \textit{mht} algorithm, which reconstructs $E\mathrm{_T^{miss}}$ as the negative of the transverse momentum ($p_{\mathrm{T}}$) vector sum of all jets reconstructed from calorimeter topological clusters at the HLT, was used as the default reconstruction algorithm. In 2017, the \textit{PUFit} algorithm which calculates the $E\mathrm{_T^{miss}}$ as the negative $p_{\mathrm{T}}$ sum of all calorimeter topological clusters corrected for pile-up has been chosen as the default algorithm given its better performance in high pile-up regions. Figure~\ref{Fig:met_rate} shows a comparison of the trigger cross-section as a function of the pile-up for both algorithms and Figure~\ref{Fig:met_eff} show the efficiencies of the \textit{mht} and \textit{PUFit} algorithms versus $E\mathrm{_T^{miss}}$ (left) and pile-up (right).

\section{Summary}
The LHC conditions and performance pose a continuous challenge to the ATLAS trigger system throughout the ongoing Run-2. To maintain the efficient and reliable event selections of the trigger, many improvements in algorithm performance and robustness across all signatures have been made to deal with increasing trigger rates due to the increase in luminosity and pile-up in 2017. Only a few selected improvements were highlighted here, however the changes made in the various signatures have already shown great improvements and performance. Further improvements to the ATLAS trigger system in terms of hardware additions are planned for with the commissioning of FTK.

\end{document}